\documentclass[letter]{aa}
\usepackage{amsmath,graphicx,amssymb}
\usepackage{txfonts}
\usepackage{natbib}
\usepackage{hyperref}

\newcommand{\Teff}{$T_\mathrm{eff}$}
\newcommand{\logg}{$\log g$}
\usepackage{color}

\begin{document}

\title{The astrophysical parameters of chemically peculiar stars from automatic methods}

\author{E. Paunzen\inst{1}}
\institute{Department of Theoretical Physics and Astrophysics, Masaryk University,
Kotl\'a\v{r}sk\'a 2, 611\,37 Brno, Czech Republic \\
\email{epaunzen@physics.muni.cz}
}

\date{}

\abstract
{The chemically peculiar (CP) stars of the upper main sequence are excellent astrophysical laboratories for investigating the diffusion, 
mass loss, rotational mixing, and pulsation in the presence and absence of a stable local magnetic field. For this, we need a homogeneous
set of parameters, such as effective temperature (\Teff) and surface gravity (\logg), to locate the stars in the Hertzsprung-Russell diagram so that we can then estimate the mass, radius, and age.}
{In recent years, the results of several automatic pipelines have been published; these use various techniques and data sets, including \Teff\ and \logg\ values for 
millions of stars. Because CP stars are known to have flux anomalies, these astrophysical parameters must be tested for their
reliability and usefulness. If the outcome is positive, these can be used to analyse the new and faint CP stars published recently.}
{I compared published \Teff\ and \logg\ values of a set of CP stars, which are mostly based on high-resolution spectroscopy,
with values from four automatic pipeline approaches. In doing so, I searched for possible correlations and offsets.}
{I present a detailed statistical analysis of a comparison between the `standard' and published  \Teff\ and \logg\ values. The
accuracy depends on the presence of a magnetic field and the spectral type of the CP subgroups. However, I obtain standard deviations of between 2\% and
20\%.}
{Considering the statistical errors, the astrophysical parameters from the literature can be used for CP stars, although caution is advised for magnetic CP stars.}


\titlerunning{Astophysical parameters of CP stars}
\authorrunning{E. Paunzen}

\maketitle

\section{Introduction} \label{introduction}

The chemically peculiar (CP) stars of the upper main sequence
have been targets of astrophysical study since their discovery by
the American astronomer Antonia Maury (1897). Most of the early
research was devoted to detecting peculiar features in their spectra
and to characterising their photometric behaviour. 

According to \citet{1974ARA&A..12..257P}, CP stars are commonly subdivided
into four classes: metallic line (or Am) stars (CP1), magnetic Ap
stars (CP2), HgMn stars (CP3), and He-weak stars (CP4). The CP1 stars are A- and early F-type objects and are defined by the 
discrepancies found in the spectral types derived from the strengths of the
\ion{Ca}{II} K line and the hydrogen and metallic lines. 
In comparison to the spectral types derived from the hydrogen lines, the \ion{Ca}{II}
K-line types appear too early, and the metallic-line types too late.
CP1 stars do not show strong, global magnetic fields \citep{2010A&A...523A..40A}
and are characterised by underabundances of calcium
and scandium and overabundances of the iron peak and heavier
elements. CP1 stars are primarily members of binary systems with
orbital periods in the range between 2 and 10 days, and their rotational
velocities are believed to have been reduced by tidal interactions,
which has enabled diffusion to act \citep{2009AJ....138...28A}.
The observed abundance pattern of CP1 stars is defined by the
diffusion of elements and the disappearance of the outer convection
zone associated with helium ionisation because of the gravitational
settling of helium \citep{2005A&A...443..627T}. These latter authors predict a
cut-off rotational velocity for such objects (about 100 km s$^{-1}$),
above which meridional circulation leads to a mixing in the stellar
atmosphere.

The CP2 stars are distinguished by their strong, globally organised magnetic 
fields that range up to several tens of kG \citep{2021A&A...652A..31B}. 
In CP2 (and CP4) stars, due to additional magnetic diffusion, the
chemical abundance concentrations at the magnetic poles, as well as
the spectral and related photometric variabilities, are also easily
understood, as are the radial velocity variations of the appearing and
receding patches on the stellar surface \citep{2015MNRAS.454.3143A}.
These inhomogeneities are
responsible for the strictly periodic changes observed in the spectra 
and brightness of many CP2 stars, which are explained by the
oblique rotator model \citep{1950MNRAS.110..395S}. Therefore, the observed periodicity
of variation is the rotational period of the star.

\begin{table*}[t]
\begin{center}
\caption{Coefficients of the corrections $\Delta$\Teff\ and $\Delta$\logg\ of the four investigated references. The meaning is
$T_\mathrm{eff}^\mathrm{corr} = a + T_\mathrm{eff}^\mathrm{publ} + (bT_\mathrm{eff}^\mathrm{publ}$) and
$\log g^\mathrm{corr} = a + \log g^\mathrm{publ} + (b\log g^\mathrm{publ}$). The values by \citet[][]{2019AJ....158..138S} 
could not be checked for the \logg\ of CP4 stars because they include an insufficient number of objects. 
If no values $a$ and $b$ are listed, the published astrophysical parameters can be used
as they are. The quantity $\sigma$ gives the standard deviation of the calibrated parameter from
the `standard value' from  \citet{2018MNRAS.480.2953G,2019MNRAS.487.5922G} in per cent. 
The errors in the final digits of the corresponding quantity are given in parentheses.}.
\label{coefficients}
\begin{tabular}{lccc|ccc|c}
  \hline
 & \multicolumn{3}{c|}{$\Delta$\Teff} & \multicolumn{3}{c|}{$\Delta$\logg} & \\
 & $a$ & $b$ & $\sigma$ & $a$ & $b$ & $\sigma$ & N\\
 & & & (\%) & & & (\%) & \\
\hline
\citet[][]{2019AandA...628A..94A,2022AandA...658A..91A} & & & & & & & \\
CP1     & +896(616)     & $-$0.123(77) & 3.6 & +1.46(50) &      $-$0.377(126) &       4.3 & 54 \\
CP2     & & &                   12.7    & & &                           8.9 & 83 \\
CP3     & +8554(1276) & $-$0.664(114) & 8.8     & 2.45(54) &    $-$0.623(136) & 4.9 & 49 \\
CP4     & & &                   17.4    & & &                           10.8 & 8 \\
\hline
\citet[][]{2019AJ....158..138S}  & & & & & & & \\
CP1 & & &                               2.2     & +1.53(25) &   $-$0.387(63) &       3.8 & 102 \\
CP2     & +1698(496) &  $-$0.157(48) &  9.2     & +3.77(61) &   $-$0.953(156) &       8.9 & 125 \\
CP3     & +7986(725) &  $-$0.617(62) &  8.6     & +2.50(68) &   $-$0.633(173) &       4.8 & 86 \\
CP4     & +10377(1426) &        $-$0.695(90) &  10.8 & & & & 10 \\      
\hline
\citet[][]{2023AandA...674A..28F} & & & & & & & \\
CP1 ($T_\mathrm{eff}^\mathrm{publ} <$\,8500\,K) & & &                           2.2     & +2.30(43) &     $-$0.569(110) & 4.5 & 58 \\
CP2     & & &                           11.7 &  +3.32(79) &     $-$0.832(202) &       15.5 & 71 \\
CP3     & +5491(1154) & $-$0.411(97) &  8.3     & +3.00(35) &   $-$0.764(92)& 6.2 & 65 \\
CP4 & +2707 & &                 12.5    &       & &                     8.6 & 9 \\
\hline
\citet{2023MNRAS.524.1855Z} & & & & & & & \\
CP1     & +4283(288) &  $-$0.552(35) &  4.5     & +2.79(15) &   $-$0.705(39) &       4.1 & 74 \\
CP2     & +5613(237) &  $-$0.659(17) &  7.7     & +3.19(16) &   $-$0.773(46) &       8.8 & 93 \\
CP3     & +10592(619) & $-$0.891(36) &  9.9     & +3.41(14) &   $-$0.842(42) &       5.4 & 63 \\
CP4     & +11990(2531) &        $-$0.768(132) & 8.2 &   +3.69(16) &     $-$0.899(52) &       4.6 & 13 \\
\hline
\end{tabular}
\end{center}
\end{table*}

The CP3 stars are characterised by strong lines of ionised Hg and/or Mn
with overabundances by up to six orders of magnitude relative to their
solar abundances \citep{2018MNRAS.480.2953G}. Several mechanisms play
significant roles in our understanding of these extreme peculiarities:
radiatively driven diffusion, mass loss, mixing, light-induced drift,
and possibly weak magnetic fields. However, no satisfactory model exists 
to explain the abundance pattern \citep{2003A&A...397..267A}.

The CP4 stars are the hottest CP objects up to
early B-types, where the mass-loss and stellar winds become significant
\citep{2007A&A...468..263C}. Initially, the CP4 stars were defined as He-weak stars only.
Later on, it was proposed that this class also includes He-strong stars \citep{1977A&AS...30...11P}.
However, the latter are rare and are not included in the present analysis.
The He lines of these objects are anomalously weak or strong for their
spectral type (effective temperature).
The shape of the Balmer continuum of CP4 stars differs from 
that predicted by models with standard solar helium abundances. Also, these stars display 
H$\alpha$ emission (especially He-strong stars) and spectroscopic and photometric variability \citep{1984A&AS...55..259N}, as well as variations in line intensities, radial velocities,
luminosity, colour, and magnetic field
strength \citep{1977A&AS...30...11P}.

Many new and relatively faint CP stars have been discovered 
\citep{2019ApJS..242...13Q,2020A&A...640A..40H,2021A&A...645A..34P,2022ApJS..259...63S} thanks to the new spectroscopic data from the $Gaia$ satellite \citep{2023A&A...674A..27A}
and the Large Sky Area Multi-Object Fibre Spectroscopic Telescope \citep[][LAMOST]{2012RAA....12.1197C}.
For further statistical analysis, we need the astrophysical parameters of our target stars 
(\Teff, \logg\ or luminosity, and mass). In order to be able to draw robust conclusions, it is most important that we obtain homogeneity in these parameters. 
\citet{2008A&A...491..545N} showed that due to the abnormal colours, determining \Teff\ for CP stars 
using photometry is not straightforward. These authors presented a comprehensive study of 
the three main photometric systems (Johnson, Geneva, and Str{\"o}mgren-Crawford) together with
a new relation for bolometric correction. However, such photometric data are unavailable for the newly
discovered CP stars.

A way out of this dilemma is to use astrophysical parameters automatically determined by 
pipeline software based on various photometric and spectroscopic data. This Letter presents 
a statistical analysis comparing the \Teff\ and \logg\ from high-resolution spectroscopy and
four automatic methods. I searched for offsets and calculated corrections in order to improve the published
astrophysical parameters for all subgroups of CP stars.

\section{Target selection and used calibrations} \label{target_selection}

The CP stars published by 
\citet{2018MNRAS.480.2953G,2019MNRAS.487.5922G} were taken to test the astrophysical parameters. 
These authors compiled well-established objects
with abundances deduced from high-resolution spectroscopic observations. Therefore, most of the
\Teff\ and \logg\ values are also based on these spectra, making them independent of any 
photometric calibrations. For further analysis, 96 CP1, 133 CP2, 87 CP3, and 18 CP4 stars were extracted.
The following approaches were used to determine the astrophysical
parameters of the target stars, and the corresponding lists were matched using either coordinates or $Gaia$ identifications.

{\it \citet[][StarHorse2021]{2019AandA...628A..94A,2022AandA...658A..91A}}: These authors combined parallaxes 
and photometry from the $Gaia$ DR3 together with the photometric catalogues
of Pan-STARRS 1 \citep{2013ApJS..205...20M}, 2MASS \citep{2006AJ....131.1163S}, AllWISE \citep{2013yCat.2328....0C}, 
and the SkyMapper DR2 \citet[without the $u$ filter; ][]{2019PASA...36...33O} to derive
Bayesian stellar parameters, distances, and extinctions using the StarHorse code \citep{2018MNRAS.476.2556Q}. This latter is a Bayesian parameter-estimation code that compares many observed quantities to stellar evolutionary
models. Given the set of observations, plus several priors, it finds the posterior probability over a
grid of stellar models, distances, and extinctions. {\it \citet[][]{2019AandA...628A..94A,2022AandA...658A..91A}} concluded that the systematic
errors of the astrophysical parameters are smaller than the nominal uncertainties for most objects.

{\it \citet[][The Revised TESS Input Catalog]{2019AJ....158..138S}}: The procedure used by these authors is based on the 
apparent magnitude in the $TESS$ bandpass ($T$), taking into account the stellar evolutionary phases.
They used PHOENIX model atmospheres together with photometric data and calibrations from the 
$Gaia$ DR2 and 2MASS catalogues. All calibrations are listed in \citet{2018AJ....156..102S}.

{\it \citet[][Gaia DR3 Apsis]{2023AandA...674A..28F}}: This is the pipeline software developed by the $Gaia$ consortium.
These authors analysed astrometry, photometry, BP/RP, and RVS spectra for objects across the Hertzsprung-Russell diagram (HRD). 
Their method was compared and validated with star cluster data, asteroseismological results, and several other
references.

{\it \citet{2023MNRAS.524.1855Z}}: These authors used $Gaia$ DR3 XP spectra to derive astrophysical parameters (\Teff, \logg, and
[Fe/H]) together with extinction values and corrected parallaxes. They applied a machine-learning model to 
directly predict stellar parameters from XP spectra 
with a training set from a model
of stellar atmospheric parameters from the LAMOST survey.  
This approach is superior because it models all relevant parameters significantly affecting the observed spectra. 
To this end, the authors used 2MASS and WISE photometry.

\begin{figure}
\begin{center}
\includegraphics[width=0.48\textwidth]{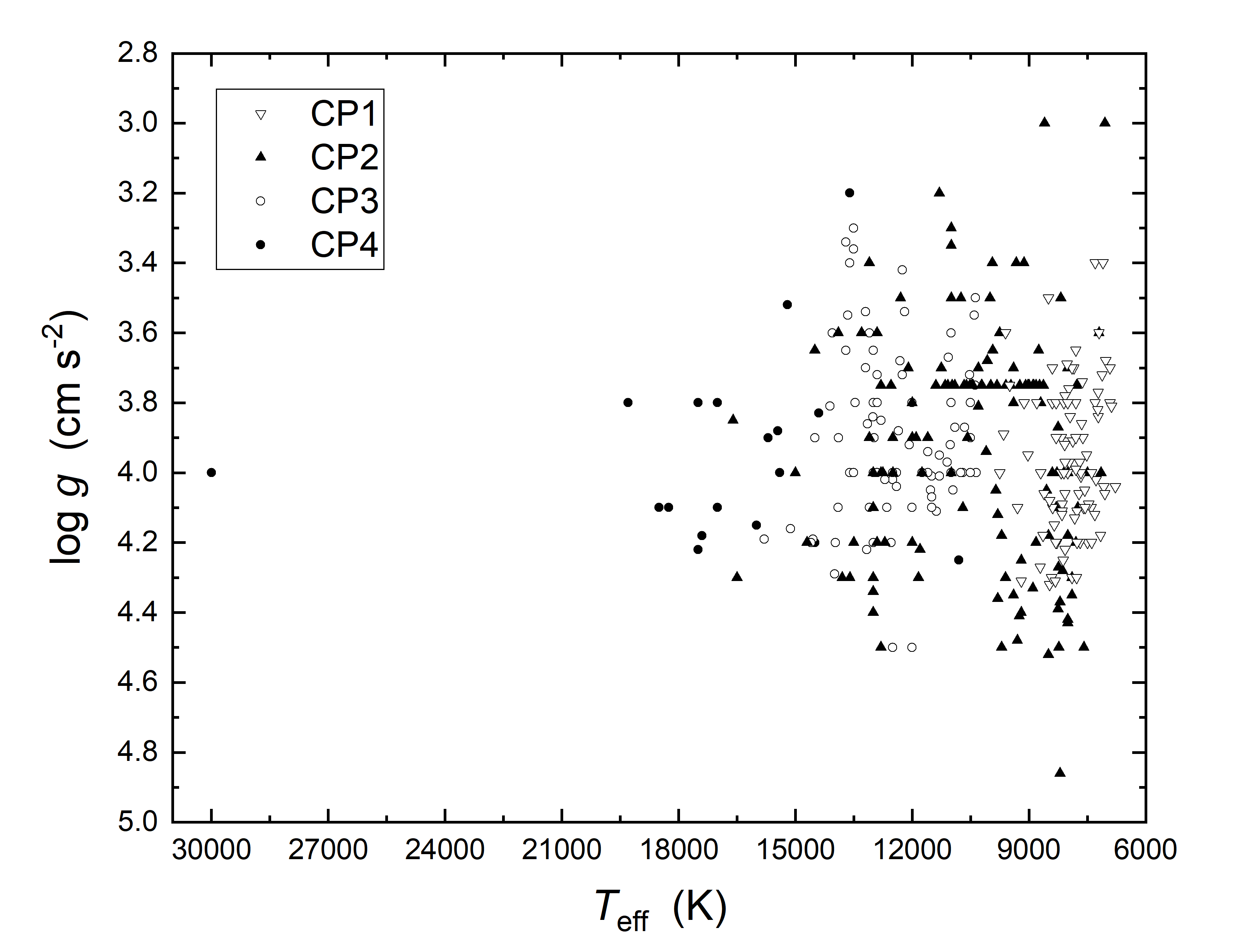}
\caption{Hertzsprung-Russell diagram of the target sample. The members of the four CP subgroups are taken from
\citet{2018MNRAS.480.2953G,2019MNRAS.487.5922G}.}
\label{HRD}
\end{center}
\end{figure}

\begin{figure}
\begin{center}
\includegraphics[width=0.48\textwidth]{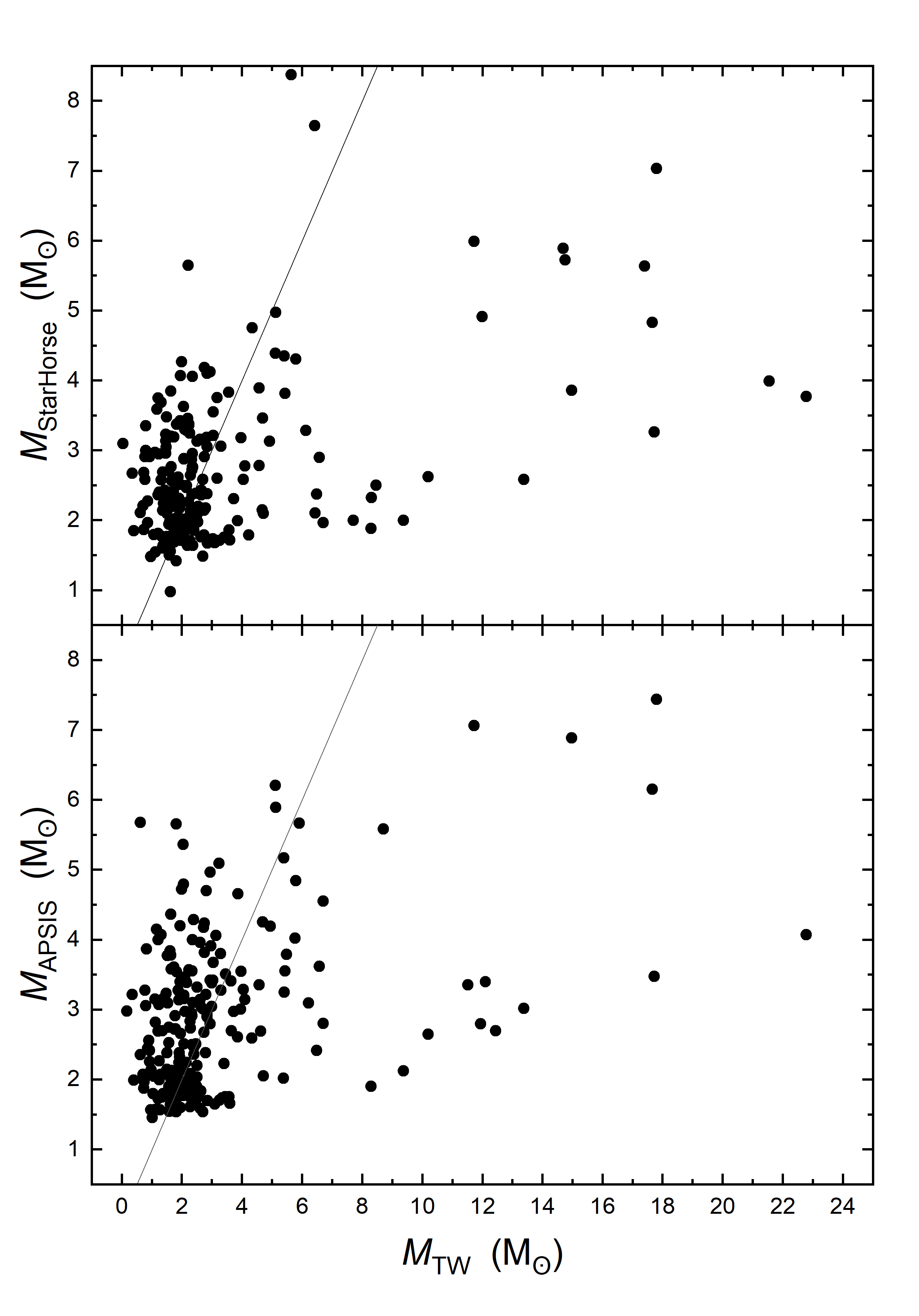}
\caption{Comparison of the masses calibrated using the `standard' values from \citet{2018MNRAS.480.2953G,2019MNRAS.487.5922G}
and the ones published by \citet[][upper panel]{2022AandA...658A..91A} and \citet[][lower panel]{2023AandA...674A..28F}. No clear
correlation is visible, and there are many outliers. The sample was not divided into the four subgroups. The abbreviation `TW' means
`this work'.}
\label{masses}
\end{center}
\end{figure}

\begin{table*}
\begin{center}
\caption{Ten CP2 stars with exceptionally high masses for their \Teff\ and \logg\ values. Further investigation is necessary 
to find the reasons for these apparent anomalies.}
\label{outliers}
\begin{tabular}{lcccccccc}
\hline
Name & \Teff & \logg & $D$ & $A_\mathrm{V}$ & $M_\mathrm{V}$ & B.C. & $\log(L/\mathrm{L}_\odot)$ & $M$ \\
& (K) & (cm\,s$^{-2}$) & (pc) & (mag) & (mag) & (mag) & & (M$_\odot$) \\
\hline
HD 42075        &       7590    &       4.50    &       277     &       0.03    &       1.75    &       0.00    &       1.20    &       6.12    \\
HD 42605        &       8250    &       4.39    &       281     &       0.34    &       1.37    &       0.03    &       1.34    &       4.71    \\
HD 43901        &       8000    &       4.18    &       347     &       0.16    &       0.36    &       0.03    &       1.74    &       8.30    \\
HD 52847        &       8200    &       4.86    &       287     &       0.02    &       0.86    &       0.03    &       1.55    &       22.77   \\
HD 55540        &       8230    &       4.50    &       518     &       0.09    &       0.75    &       0.03    &       1.59    &       10.78   \\
HD 62244        &       8550    &       4.05    &       409     &       0.03    &       0.22    &       0.02    &       1.81    &       5.43    \\
HD 91087        &       8500    &       4.52    &       394     &       0.25    &       1.27    &       0.02    &       1.38    &       6.21    \\
HD 97394        &       8000    &       4.43    &       379     &       0.04    &       0.86    &       0.03    &       1.55    &       9.37    \\
HD 102333       &       8240    &       4.27    &       467     &       0.17    &       0.54    &       0.03    &       1.67    &       7.70    \\
HD 110274       &       8135    &       4.28    &       365     &       0.41    &       1.16    &       0.03    &       1.42    &       4.68    \\
\hline
\end{tabular}
\end{center}
\end{table*}

\section{Results and conclusions} \label{results_conclusions}

I calculated the differences (`standard' minus literature value; $\Delta$\Teff\ and $\Delta$\logg) for each CP subgroup and 
reference for \Teff\ and \logg, respectively. I then searched for correlations using these differences.
Table \ref{coefficients} presents the results of this statistical analysis. 

As can be seen from the last column, the number of available data points varies because the subgroups have
different sizes, but also because the hotter stars (CP3 and CP4) are, in general, more difficult to
calibrate and are often missing in the automatic analysis. The astrophysical parameters for the CP1 and CP4 stars by \citet[][]{2023AandA...674A..28F} are limited in \Teff\
as listed in Table \ref{coefficients}. The \logg\ values of CP4 stars listed in \citet[][]{2019AJ....158..138S}
cannot be checked because an insufficient amount of data is available for this class of objects.  

No correlations were found between $\Delta$\Teff\ and \logg\ and between $\Delta$\logg\ and \Teff. 
For some combinations, the published values can be used as they are; for example, the values for CP2
stars by \citet{2019AJ....158..138S}. As a quality indicator,
I calculated the quantity $\sigma$, which gives the standard deviation of the calibrated parameter from
the `standard value' from  \citet{2018MNRAS.480.2953G,2019MNRAS.487.5922G} in per cent.
In general, $\sigma$ is the smallest for the CP1 and CP3 subgroups, which are non-magnetic. 

I calculated the mean $\sigma$  of the individual calibrated \Teff\ and \logg\ values for all available data, which
includes the entire sample of 334 stars, obtaining the following $\sigma$ values for the
four subgroups (CP1-4): [3.3,4.5], [9.5,8.1], [8.6,5.7], and [12.4,7.3].

I then compared the published masses from \citet{2022AandA...658A..91A} and \citet{2023AandA...674A..28F}
with those estimated from the `standard' values by \citet{2018MNRAS.480.2953G,2019MNRAS.487.5922G}. The latter publication does not include
them, and so I estimated the masses using the luminosity and the formula
\begin{equation}
\log(M/\mathrm{M}_\odot) = \log(g/g_\odot) - 4\log(T_\mathrm{eff}/T_\mathrm{eff\,\odot}) + \log(L/\mathrm{L}_\odot)    
,\end{equation}
with the recommended IAU values for the Sun ($\log g_\odot$ = 4.438 and $T_\mathrm{eff\,\odot}$ = 5772\,K).
To calculate the individual masses, I used the extinctions from \citep{reddening_paper} and
\citet{2019ApJ...887...93G}. The distances were taken from \citet{2021AJ....161..147B} who used the $Gaia$ EDR3
and a prior constructed from a three-dimensional model of our Galaxy. Because most of the stars are closer than 500\,pc,
the reddening can be mostly neglected. The $V$ magnitudes were taken from \citet{2001KFNT...17..409K} and \citet{2022A&A...661A..89P}.
Finally, the bolometric corrections are those from 
\citet{2008A&A...491..545N} for the magnetic CP stars and from \citet{1994MNRAS.268..119B} for the non-magnetic CP1 and CP3
objects. 

Figure \ref{masses} shows the comparison of the different values, from which several conclusions can be made:

\begin{itemize}
    \item The masses for the CP1 stars generally agree relatively well. These are the coolest and therefore the least massive
    objects, and are those for which the pipelines are optimised. 
    \item The high-mass end (CP3 and CP4 stars) is underestimated in the literature. Most objects are not included
    because they are too hot for the pipelines  used.
    \item For ten CP2 stars (listed in Table \ref{outliers}), we find excessively high masses for their \Teff\ and \logg\ values. This is caused by the high
    luminosities, for which we have no explanation. No correlations with the known magnetic field strengths \citep{2021A&AT...32..137B}
    were found. The reddening values and bolometric corrections are not exceptional. Also, direct conversion
    of the parallaxes does not change the results. Binarity could play a role, which,  for a mass
    ratio of one, would shift the location by $\Delta M_\mathrm{V}$\,=\,0.75\,mag, but this is not sufficient to explain the observed high luminosities. From the astrometric
    measurements by $Gaia$, only three stars (HD 55540, HD 102333, and HD 110274) show hints of being binary systems \citep{2019A&A...623A..72K}.
    Additional data, such as classification resolution spectra and a new 
    homogeneous analysis, are needed to shed more light on the nature of these objects and the possible sources of  error. However, we can also speculate that
    the discrepancies found can be used to detect new magnetic CP stars.
\end{itemize}

I used the correlations presented in the paper by \citet{2021A&A...655A..91K} to verify the masses further. These authors derived a 
mass--effective-temperature--surface gravity relation for main sequence stars in the range of 
6400\,$<$\,\Teff\,$<$\,20000K with \logg\,$>$\,3.44, respectively. These ranges cover most of the CP star sample.
I checked the results from this calibration in comparison with those from \citet{2022AandA...658A..91A} and \citet{2023AandA...674A..28F}
for our sample. No correlation exists up to $\mathrm{2.5M}_\odot$. For larger masses, the values 
from the literature show some linear relation but are systematically too small. 

The presented analysis shows that the published astrophysical parameters, especially for the non-magnetic and cooler CP stars, 
are statistically functional. 
With this in mind, several tasks are awaiting future research projects. Many 
new, faint CP stars have been discovered in recent years for which only masses and ages have been published
by \citet{2020A&A...640A..40H}. However, the precise location of the objects in the HRD is needed to fit isochrones.
Future studies on the rotational behaviour of CP stars \citep{2021A&A...656A.125F}
will also require a well-established HRD so that statistically sound conclusions can be drawn.

\begin{acknowledgements}
This work was supported by the grant GA{\v C}R 23-07605S.
I thank Klaus Bernhard, Stefan H{\"u}mmerich, and Martin Netopil for discussing various topics and 
helping to improve the paper significantly.
This work has made use of data from the European Space Agency (ESA) mission 
{\it Gaia} (\url{https://www.cosmos.esa.int/gaia}), processed by the {\it Gaia} Data 
Processing and Analysis Consortium (DPAC, \url{https://www.cosmos.esa.int/web/gaia/dpac/consortium}). 
Funding for the DPAC has been provided by national institutions, in particular, the institutions
participating in the {\it Gaia} Multilateral Agreement. This research has made use of the SIMBAD database,
operated at CDS, Strasbourg, France.
\end{acknowledgements}

\bibliographystyle{aa}
\bibliography{aa48086_23}

\end{document}